\documentclass[11pt]{article}

\usepackage{amsfonts}
\usepackage{graphicx}
\usepackage{amsmath}

\newtheorem{theorem}{Theorem}
\newtheorem{acknowledgement}[theorem]{Acknowledgement}

\input{tcilatex}
\makeatletter \@addtoreset{equation}{section}

\newcommand{\be}{\begin{equation}}
\newcommand{\ee}{\end{equation}}
\newcommand{\bea}{\begin{eqnarray}}
\newcommand{\eea}{\end{eqnarray}}
\begin{document}

\title{ \rightline{\mbox {\normalsize
{GNPHE/0501-VACBT/0501}}} \textbf{Non-commutative ADE geometries as}\\
\textbf{holomorphic wave equations}\\
[.3cm] }
\author{Adil Belhaj$^{1,}$\thanks{%
abelh633@mathstat.uottawa.ca}\ ,\ J{\o }rgen Rasmussen$^{2,}$\thanks{%
rasmusse@crm.umontreal.ca}\ ,\ El Hassan Saidi$^{3,}$\thanks{%
esaidi@ictp.trieste.it}\ ,\ Abdellah Sebbar$^{1,}$\thanks{%
asebbar@mathstat.uottawa.ca} 
\\[.4cm]
{\small \mbox{}$^1$\ Department of Mathematics and Statistics, Ottawa
University}\\
{\small \ 585 King Edward Ave., Ottawa, ON, Canada K1N 6N5}\\
[.1cm] {\small \mbox{}$^2$\ Department of Mathematics and Statistics,
Concordia University }\\
{\small \ 7141 Sherbrooke St. W, Montr\'eal, PQ, Canada H4B 1R6}\\
[.1cm] {\small \mbox{}$^3$\ Virtual African Centre for Basic Science and
Technology, VACBT,}\\
{\small \ focal point, Lab/UFR-Physique des Hautes Energies, Facult\'{e} des
Sciences, Rabat, Morocco}}
\maketitle

\begin{abstract}
Borrowing ideas from the relation between classical and quantum mechanics,
we study a non-commutative elevation of the $ADE$ geometries involved in
building Calabi-Yau manifolds. We derive the corresponding geometric
hamiltonians and the
holomorphic wave equations representing these non-commutative geometries. 
The spectrum of the holomorphic waves is interpreted as the quantum 
moduli space.
Quantum $A_{1}$ geometry is analyzed in some details and is found to be linked
to the Whittaker differential equation.
\end{abstract}


\thispagestyle{empty}

\newpage \setcounter{page}{1} \newpage

\section{Introduction}

\qquad In recent years, there has been a great deal of interest in
non-commutative (NC) spaces in connection with string theory. Common to many
of these studies is that the non commutativity stems from the D-brane
physics in the presence of a B-field \cite{SW}. Similar NC structures have
been applied to Calabi-Yau compactifications. 
The underlying idea in this context is to express the
non commutativity in terms of discrete isometries of
orbifolds. This was successfully done in \cite{BL} for the quintic, and it has
been extended to K3 surfaces \cite{KL, BMR} and 
higher-dimensional orbifolds \cite{BS1}. The NC aspect of such
hypersurfaces is also important for the stringy resolution of
singularities as it offers an alternative to the standard
resolutions obtained by deformations of the complex or K\"{a}hler structures
of the Calabi-Yau manifolds. 

An objective of the present paper is to develop a new and essentially
non-geometric approach to
NC Calabi-Yau manifolds, based on ideas from quantum mechanics. 
This also offers a new take on the moduli
space of resolved singularities. In this paper, a moduli space is meant to
denote a space spanned by the degrees of freedom in the system.
Our focus is on $ADE$ geometries,
which we represent by certain holomorphic,
partial differential operators. Such an operator acts on the space of
holomorphic functions $\Psi$ on $\mathbb{C}^{3}$, thereby
defining a wave equation. The spectrum of wave functions solving
this equation is accordingly interpreted as the moduli space of the NC elevation
of the associated $ADE$ geometry. We consider in some details
the case $A_1$, and we find that it is linked to the Whittaker 
differential equation.

Our wave-functional approach may be applied to more general geometries
than the $ADE$ spaces. It thus offers a whole new description of NC
elevations of ordinary geometries. A more general exposition may be found
in \cite{work} while a more conventional approach to NC Calabi-Yau manifolds 
may be found in \cite{ss,Be}.

The present paper is organized as follows: In Section 2, we outline how
our NC elevations of ordinary geometries mimic the 
quantization of classical mechanics in the hamiltonian formalism.
Section 3 concerns the quantization of the $ADE$ geometries,
while the wave-equation representations of the resulting NC
geometries are discussed in Section 4. Section 5 contains
some concluding remarks.

\section{Basic correspondence}

Our construction of NC $ADE$ geometries as elevations of ordinary, 
commutative $ADE$ geometries is based on an extension of the 
relation between classical and quantum mechanics. 
We are thus exploring a similarity 
between commutative $ADE$ geometries and classical mechanics
on one hand, and NC $ADE$ geometries and quantum mechanics on the other
hand. The basic ideas are outlined in the following.

\subsection{Ordinary $ADE$ geometries  and classical mechanics}

A hypersurface in the three-dimensional complex space 
$\mathbb{C}^3$ generated by $(z^1,z^2,z^3)=(x,y,z)$ may be described
by an algebraic equation of the form
\begin{equation}
  V(x,y,z)=\epsilon.
\label{Veps}
\end{equation}
The explicit (polynomial) potentials $V(x,y,z)$ of our interest will be 
specified below. The parameter $\epsilon$ is independent of the complex 
coordinates $(x,y,z)$, and
may be seen as parameterizing the family or orbit of hypersurfaces
characterized by a given potential $V$. It is observed that a point
$(x,y,z)\in\mathbb{C}^3$ can lie on at most one hypersurface in a
given orbit. 
Since $\epsilon$ is constant, 
a point $(x_0,y_0,z_0)$ on the hypersurface (\ref{Veps})
is a singular point if the gradient 
$\nabla V=(\partial_xV,\partial_yV,\partial_zV)$ vanishes at that point:
\begin{equation}
  \nabla V(x_0,y_0,z_0)=(0,0,0).
\label{nablaV}
\end{equation}
It should be evident that the above extends to hypersurfaces in 
the higher-dimensional complex spaces $\mathbb{C}^n$.
The $ADE$ geometries of our interest are all singular at exactly
one point. The associated
potentials will be chosen such that the $ADE$ geometries correspond
to $\epsilon=0$ in their respective orbits, and such that 
the singularities appear at the origin $(x_0,y_0,z_0)=(0,0,0)$. 

The hamiltonian description of classical mechanics is quite analogous.
One may be interested in characterizing the configurations corresponding
to a fixed energy $E$. This amounts to solving the equation
\begin{equation}
  \mathcal{H}(q_1,\ldots,q_n,p_1,\ldots,p_n)=E
\label{HE}
\end{equation}
where $\mathcal{H}$ denotes the hamiltonian.
The solutions define a hypersurface in the 
$2n$-dimensional phase space. Up to some well-known signs, 
Hamilton's equations express the
time derivatives of the canonical coordinates $q_j,p_j$, $j=1,\ldots,n$, 
in terms of the gradient of the hamiltonian. A singular point of the 
fixed-energy hypersurface thus corresponds to the simultaneous 
vanishing of all these time derivatives.

With this analogy we are thus considering a similarity between 
the orbit of hypersurfaces
based on $V$ and the physical system described by $\mathcal{H}$.
The individual hypersurfaces characterized by $\epsilon$ play the roles
of specific energy levels given by $E$. Singularities appear
where the associated gradients vanish.

There are of course important differences and further similarities
between these two scenarios. Here, though, we will not be concerned with them.
Rather, our objective is to explore the consequences of mimicking 
the quantization of classical mechanics in the realm of hypersurfaces
of the form (\ref{Veps}).

\subsection{NC $ADE$ geometries  and quantum mechanics}

We will consider a combination of the Heisenberg and Schr\"odinger
pictures of quantization. As part of our construction we thus replace the 
complex coordinates $z^j$ by holomorphic operators $Z^j$ in analogy with the 
promotion of the canonical variables to quantum operators. We also 
borrow the idea of promoting the classical hamiltonian to a differential
operator acting on a space of wave functions where the eigenfunctions
correspond to the stationary states, whereas the eigenvalues represent
the allowed energy levels. In our case, the potential $V$ is replaced by
a differential operator whose eigenfunctions will
be certain holomorphic functions, while the eigenvalues will label
the NC geometries we may represent in this picture.

The geometric analogue of the Schr\"odinger wave equation which we
will discuss reads
\begin{equation}
  V(X,Y,Z)\Psi=\epsilon\Psi.
\label{VPsi}
\end{equation}
We naturally require that this reduces to (\ref{Veps}) in the
classical (commutative) limit. We may decompose
the `quantum potential' $V(X,Y,Z)$ as
\begin{equation}
  V(X,Y,Z)=H+V(x,y,z)
\label{VHV}
\end{equation}
where the holomorphic (partial) differential operator $H$ vanishes
in the classical limit:
\begin{equation}
  H\rightarrow0.
\label{class}
\end{equation}
The `geometric hamiltonian' $H$ is constructed by replacing the
coordinates $z^j$ by a differential-operator realization of the
NC coordinates $Z^j$ subject to a normal-ordering procedure
to be discussed below. This also justifies the use of the same
symbol $V$ to denote the quantum potential as in the classical case.
Keeping the decomposition (\ref{VHV}) and the
classical limit (\ref{class}) in mind, the partial
differential operators $Z^j$ may be viewed as NC perturbations
of the original commutative coordinates $z^j$ where the perturbative
terms vanish in the classical limit.

We will derive and study differential equations of the form
\begin{equation}
  H(x,y,z;\partial_x,\partial_y,\partial_z)\Psi(x,y,z)=
   \left(\epsilon-V(x,y,z)\right)\Psi(x,y,z),
\label{HPsiVPsi}
\end{equation}
where the $ADE$ geometries correspond to $\epsilon=0$.
Deriving these equations essentially amounts to devising 
an appropriate normal ordering of the NC coordinates.
This is discussed in the following section.
Working out the corresponding quantum moduli space
amounts to finding the spectrum of eigenfunctions $\Psi$
in (\ref{HPsiVPsi}). This is a highly non-trivial task as it requires
solving complicated partial differential equations.
The general solution is beyond the scope of the
present work, though we will present the solution in the case of $A_1$.

In brief, our construction offers a novel and essentially 
non-geometric representation of NC $ADE$
geometries as holomorphic wave equations on $\mathbb{C}^{3}$.
The associated moduli spaces of the NC $ADE$ geometries
are given in terms of the spectrum of holomorphic waves solving these
differential equations.

It is well known that a singular $ADE$ geometry may be deformed by
adding a polynomial term $f(x,y,z)$ to the defining potential $V(x,y,z)$,
where either $f(0,0,0)\neq(0,0,0)$ or $\nabla f(0,0,0)\neq0$.
The corresponding NC elevation is represented by a wave equation
of the form
\begin{equation}
  H(x,y,z;\partial_x,\partial_y,\partial_z)\Psi(x,y,z)=
   -(V(x,y,z)+f(x,y,z))\Psi(x,y,z).
\end{equation}
NC elevations of such deformations will not be considered further here.

\section{Quantization of $ADE$ geometries}

For the sake of simplicity we will limit our analysis to the complex
K3 surfaces being an
important example of compact Calabi-Yau manifolds. These surfaces 
play a crucial role
in the study of type II superstrings and in the geometric engineering of 
quantum field theories
embedded in superstring theory \cite{KKV,KMV,BFS,BS2}. A K3 surface
can have singularities corresponding to contracting
two-spheres. The intersection matrix of these two-spheres is then given by the
Cartan matrix of a Lie algebra, and one may naturally 
distinguish between three types of
singularities \cite{ABS}, namely (\textbf{a}) \textsl{ordinary singularities}
classified by the ordinary $ADE$ Lie algebras, (\textbf{b}) \textsl{affine
singularities} classified by the affine $\widehat{ADE}$ 
Kac-Moody algebras, and 
(\textbf{c}) \textsl{indefinite singularities} classified by the indefinite Lie algebras.

Here we focus on ordinary $ADE$ singularities, while a similar analysis is 
possible for the affine extensions. Near such an ordinary singularity, a 
K3 surface may be viewed as an ALE space defined by an orbifold structure 
$\mathbb{C}^{2}/G$, where $G$
is a discrete group depending on the $ADE$ singularity in question. These
orbifolds can be expressed as hypersurfaces in $\mathbb{C}^{3}$ (\ref{Veps})
where $\epsilon=0$,
\begin{equation}
  V(x,y,z)=0,
\label{ADE0}
\end{equation}
and with potentials given by \cite{HOV}\footnote{We have chosen to represent
the $A$-series by $x^2+y^2+z^n$ instead of $uv+z^n$. The two representations
are related by the invertible transformation $(x,y)\rightarrow(u,v)=(x+iy,x-iy)$.
As will become clear, the choice $(x,y)$ renders the quantization straightforward.}
\begin{eqnarray}
 A_{n-1}:&&\ \ V_{A_{n-1}}(x,y,z):=x^2+y^2+z^{n},  \notag \\ 
 D_{n}:&&\ \ V_{D_{n}}(x,y,z):=x^{2}+y^{2}z+z^{n-1},  \notag \\
 E_{6}:&&\ \ V_{E_{6}}(x,y,z):=x^{2}+y^{3}+z^{4},  \nonumber \\
 E_{7}:&&\ \ V_{E_{7}}(x,y,z):=x^{2}+y^{3}+yz^{3},  \notag \\
 E_{8}:&&\ \ V_{E_{8}}(x,y,z):=x^{2}+y^{3}+z^{5}. 
\label{ADE}
\end{eqnarray}
The indices indicate the ranks of the Lie algebras. 
As already mentioned, these hypersurfaces have singularities
at the origin of $\mathbb{C}^{3}$. It is also well known that the singularity
of any one of these hypersurfaces may be
resolved in two ways, either by deforming the complex structure of the
surface (changing the shape), or by varying its K\"{a}hler structure
(changing the size). Here we are not interested in such deformations or
resolutions, but rather in a NC elevation of the $ADE$
spaces defined by the polynomial constraint equations (\ref{ADE0},\ref{ADE}). 
Following the previous section, we will
introduce a quantization procedure in which these NC $ADE$ spaces are
constructed by imposing polynomial constraints similar to 
(\ref{ADE0},\ref{ADE}) on a NC generalization of $\mathbb{C}^{3}$.

It is noted that one may also consider K3 surfaces with singularities
described by the $BCFG$ Lie algebras. The potentials defining these complex
surfaces are in general multiple-valued functions \cite{BFS,BS2}, unlike the
$ADE$ cases in (\ref{ADE}), and will not be discussed here.

We now turn to the construction of the NC embedding space. The
ordinary $n$-dimensional complex space $\mathbb{C}^{n}$ is parameterized by
the $n$ complex variables $z^{j}$, $j=1,\dots ,n$. We will parameterize its
NC counterpart $\mathbb{C}_{\Theta}^{n}$ by $Z^{j}$, $j=1,\dots ,n$,
satisfying
\begin{equation}
  \lbrack Z^{j},Z^{k}]=2\Theta^{jk}, 
\label{ZZ}
\end{equation}
where $\Theta^{jk}$ is an anti-symmetric complex tensor. 
We wish to add some comments on this non commutativity.
The first comment concerns the fact that all anti-symmetric
matrices of odd dimension are singular, i.e., not invertible.
A real hamiltonian system, on the other hand, is always
even-dimensional since each position variable is
accompanied by a conjugate momentum variable.
The singular property of $\Theta$ thus restricts the
way the operators $Z^{j}$ may be expressed in terms of 
`phase-space' variables, see (\ref{ZZP}).

The second comment concerns the complex nature of the structure constants 
$\Theta^{jk}$. They can be viewed as a complexification of the real
Seiberg-Witten parameters known to be related to the NS-NS B-field in the
description of real Moyal space \cite{SW}. In type IIB superstring theory this
complexity may have its origin in terms of a complexified K\"ahler form
$\mathcal{J}=J_{K}+iB_{NS}$ or in terms of a complex combination of the
two B-fields (RR and NS-NS), i.e., $B=B_{NS}+iB_{R}$. 

The third comment is
that the parameters $\Theta^{jk}$ play a role similar to (the normalized)
Planck's constant $\hbar$ appearing in the Heisenberg commutation relations
of non-relativistic quantum mechanics:
\begin{equation}
  \lbrack \mathcal{P},\mathcal{X}]=-i\hbar.
\end{equation}
Here $\mathcal{X}$ and $\mathcal{P}$ are the usual position and
momentum operators, respectively. 
As is well known, they admit a representation in which $\mathcal{X}=x$ 
and $\mathcal{P}=-i\hbar \partial_{x}$. 
Motivated by this, we wish to realize the NC coordinates 
$Z^j$ (satisfying (\ref{ZZ})) in terms of a linear combination of $2n$
`phase-space' variables, $\mathcal{Z}^j$ and $\mathcal{P}_j$, as follows
\begin{equation}
  Z^j=\mathcal{Z}^j+\sum_{k=1}^{n}\Theta^{jk}\mathcal{P}_k  
\label{ZZP}
\end{equation}
where $\mathcal{P}_j$ and $\mathcal{Z}^k$ are quantum operators satisfying
\begin{equation}
 [\mathcal{P}_j,\mathcal{Z}^k]=\delta_j^k,
  \qquad [\mathcal{Z}^j,\mathcal{Z}^k]=[\mathcal{P}_j,\mathcal{P}_k]=0.  
\label{PZdelta}
\end{equation}
Representing the variables $(\mathcal{Z}^j,\mathcal{P}_j)$ as 
$(z^j,\frac{\partial }{\partial z^j})$, we may represent the NC 
coordinates, $Z^j$, as first-order differential operators:
\begin{equation}
  Z^j=z^j+\sum_{k=1}^{n}\Theta^{jk}\partial_{k},\qquad \ \ \ \ \ 
   \partial_k=\frac{\partial}{\partial z^k}.  
\label{Zdiff}
\end{equation}
In this representation, the NC coordinates are thus seen to act 
as non-trivial,
holomorphic, partial differential operators on the space of holomorphic
functions $\Psi(z_1,\ldots,z_n)$ on $\mathbb{C}^{n}$, and we are one
step closer to the geometric analogue of the Schr\"odinger equation
discussed above. 

For invertible $\Theta^{jk}$, in which case the dimension $n$ must be even, 
one may introduce the `gauge potential' $A_j=\sum_{k=1}^n\Theta_{jk}z^{k}$.
The realization (\ref{Zdiff}) may then be re-expressed in terms of
the `covariant' derivative $D_j=\partial_j+A_j$ as
\begin{equation}
  D_j=\sum_{k=1}^n\Theta_{jk}Z^{k}. 
\end{equation}
This indicates that the NC elevation in these cases behaves as
switching on an external constant magnetic field 
$B^{i}=\varepsilon^{ijk}\partial_{j}A_{k}=-\varepsilon^{ijk}\Theta_{jk}$. 
Since our main interest is based on $n=3$, we will not elaborate on
this observation.

Referring to the notation in (\ref{ADE}), we will parameterize 
$\mathbb{C}_{\Theta}^{3}$ by the NC coordinates $(X,Y,Z)$ satisfying
\begin{equation}
  \lbrack X,Y]=2\alpha ,\ \ \ \ \ \ \ [Y,Z]=2\beta ,\ \ \ \ \ \ \ [Z,X]=2\gamma
\label{abc}
\end{equation}
where $\alpha,\beta,\gamma$ are (commutative)
structure constants.
It is noted that this algebra is equivalent to a central extension
of the direct sum of three $u(1)$s. That is, the $u(1)$s are originally
generated by (the commuting variables) $X,Y,Z$ while the central
element is denoted $I$. To complete the interpretation of (\ref{abc})
as this central extension, the structure constants appearing on
the right-hand sides of the commutators should all be multiplied by $I$.

Following (\ref{ZZP}) and (\ref{Zdiff}), the representations
of $(X,Y,Z)$ of our interest now read 
\begin{eqnarray}
 X &=&\mathcal{X}+\alpha \mathcal{P}_{y}-\gamma \mathcal{P}_{z}\ =\ x+\alpha
  \partial _{y}-\gamma \partial _{z},  \notag \\
 Y &=&\mathcal{Y}+\beta \mathcal{P}_{z}-\alpha \mathcal{P}_{x}\ =\ y+\beta
  \partial _{z}-\alpha \partial _{x},  \notag \\
 Z &=&\mathcal{Z}+\gamma \mathcal{P}_{x}-\beta \mathcal{P}_{y}\ =\ z+\gamma
  \partial _{x}-\beta \partial _{y}.  
\label{XYZ}
\end{eqnarray}
It is emphasized that these operators act on the local
coordinates as $[X,x]=0$, $[X,y]=\alpha $, $[X,z]=-\gamma$ etc. 
It is also noted
that one may consider various degrees of non commutativity corresponding to
\begin{eqnarray}
 \alpha &\neq &0,\qquad \beta =\gamma =0,\qquad \text{or a cyclic
  permutation,}  \notag \\
 \alpha \beta &\neq &0,\qquad \gamma =0,\qquad \qquad \text{or a cyclic
  permutation,}  \nonumber \\
 \alpha \beta \gamma &\neq &0.
\label{ca}
\end{eqnarray}
The remaining case where $\alpha=\beta=\gamma=0$ merely corresponds to 
classical geometry. Obviously, the possibility $\alpha \beta \gamma \neq 0$
has the highest degree of non commutativity.

Our next objective is to define the NC elevation of the potentials $V(x,y,z)$.
As in other `quantization' schemes, the naive substitution
\begin{equation}
  (x,y,z)\rightarrow (X,Y,Z)
\label{xyz}
\end{equation}
is ambiguous due to the simple fact that $xy=yx$ while $XY\neq YX$ 
if $\alpha\ne0$, for example, and one is faced with an ordering problem. 
According to (\ref{ADE}), we need to treat $y^{2}z$ and $yz^{3}$, 
as these monomials appear
in the $D_n$ and $E_7$ potentials, respectively. To this end, and to
put it into a more general context, we introduce the homogeneous polynomials
\begin{equation}
 M_{m}(u,v)=\sum_{j=0}^{m}a_{j}u^{j}vu^{m-j},\ \ \ \ \ \ \ \ \ \ \
  \sum_{j=0}^{m}a_{j}=1  
\label{M}
\end{equation}
of degree $m+1$ where $m$ is a non-negative integer. The arguments, $u$ and 
$v$, may be NC variables, and $M_{m}(u,v)$ is seen to reduce to the monomial 
$u^{m}v$ if $[u,v]=0$.

Now, in our case we are thus interested in
\begin{eqnarray}
 M_{m}(X,Y) &=&\sum_{j=0}^{m}a_{j}X^{j}YX^{m-j}  \notag \\
 &=&\sum_{j=0}^{m}a_{j}(\mathcal{X}+\alpha \mathcal{P}_{y} -\gamma 
  \mathcal{P}_{z})^{j}(\mathcal{Y}+\beta \mathcal{P}_{z}-\alpha 
  \mathcal{P}_{x})(\mathcal{X} +\alpha \mathcal{P}_{y}-\gamma 
  \mathcal{P}_{z})^{m-j},  
\label{MXY}
\end{eqnarray}
and we find that it may be written in the following form:
\\[0.2cm]
\textbf{Lemma}
\begin{eqnarray}
 M_{m}(X,Y) &=&\mathcal{Y}X^{m}+X^{m}(\beta \mathcal{P}_{z}-\alpha 
  \mathcal{P}_{x})+\alpha \sum_{j=0}^{m}(2j-m)a_{j}X^{m-1},  \notag \\
  M_{m}(X,X) &=&X^{m+1} . 
\label{lemma}
\end{eqnarray}
Up to commutative (hence trivial) re-arrangements (within $X^{s}$), the
ordering of the right-hand side has phase-space coordinates to the left of
the phase-space momenta. We will refer to this ordering as \emph{normal
ordering}.

Our proposal for a `natural' quantization procedure that elevates an
ordinary  hypersurface to a NC hypersurface 
now goes as follows. Let the classical hypersurface be defined by the
vanishing of a polynomial, as in the case of the $ADE$ manifolds 
(\ref{ADE0},\ref{ADE}). Since our prime goal is to construct NC elevations of
these $ADE$ spaces, we will restrict ourselves to the situation where each
monomial summand is of the form $x^{m}y$ or $z^{s}$, or similar
monomials in $\{x,y,z\}$ obtained by replacing $x$, $y$ or $z$ by one
of the other coordinates. Each of these monomials is then replaced by the
most general homogeneous polynomial in the corresponding NC coordinates 
$\{X,Y,Z\}$ (as in the first line of (\ref{MXY}), for example) satisfying that
the result of normal ordering it must itself be a homogeneous polynomial in
the phase-space variables of the NC coordinates. 
It is of course also required that the NC polynomial is properly normalized 
so that it reduces to the original
polynomial in the classical limit where $(X,Y,Z)\rightarrow (x,y,z)$. For
the class of polynomials $M_{m}(X,Y)$, this means that the right-hand side
of (\ref{lemma}) must be homogeneous in the phase-space variables, which is
ensured provided
\begin{equation}
  \sum_{j=0}^{m}(2j-m)a_{j}=0.  
\label{sum0}
\end{equation}
The symmetrized polynomial, where $a_{0}=\ldots =a_{m}=\frac{1}{m+1}$, is
seen to satisfy this condition, showing that a solution exists for all $m$.
It may appear surprising, though, that the thus defined set 
of `quantizations' of a given classical polynomial consists 
of \emph{one} polynomial only. This follows straightforwardly, 
though, from the lemma with (\ref{sum0}) imposed, since
the first part of the right-hand side of
(\ref{lemma}) is \emph{independent} of $\{a_{0},\ldots ,a_{m}\}$, and from
the fact that the quantization of $z^{s}$ is trivial. It also indicates that
our quantization procedure for a complex hypersurface like the $ADE$ spaces 
(\ref{ADE0},\ref{ADE}) results in a \emph{unique} NC hypersurface. 
In brief, the quantization
procedure replaces uniquely the classical (commutative) monomials of the
form $x^{m}y$ or $z^{s}$ by homogeneous polynomials of degree $m+1$ or $s$,
respectively, in the phase-space variables associated to the NC coordinates 
$\{X,Y,Z\}$.

Let us illustrate the uniqueness from the point of view of the relations
following from the commutative nature of the structure constants (\ref{abc}). 
Recall that $\sum_{j=0}^{m}a_{j}=1$ ensures that the NC polynomial
reduces to its commutative origin in the classical limit $(X,Y,Z)\rightarrow
(x,y,z)$, while $\sum_{j=0}^{m}(2j-m)a_{j}=0$ ensures homogeneity of the NC
counterpart of a classical monomial. For $m=1$ there is only one solution to
these two constraints: $a_{0}=a_{1}=1/2$. For $m=2$ there is the
one-parameter family of solutions
\begin{equation}
  a_{0}=a,\ \ \ \ \ \ \ a_{1}=1-2a,\ \ \ \ \ \ \ a_{2}=a,  
\label{k2}
\end{equation}
but
\begin{equation}
  ZY^{2}-2YZY+Y^{2}Z=[Z,Y]Y-Y[Z,Y]=0
\end{equation}
according to the aforementioned commutative nature of the structure
constants. Likewise for $m=3$, where
\begin{equation}
 a_{0}=b,\ \ \ \ \ a_{1}=\frac{1}{2}-2b+c,\ \ \ \ \ \ \ a_{2}=\frac{1}{2}
  +b-2c,\ \ \ \ \ \ \ a_{2}=c  
\label{k3}
\end{equation}
is the general solution, we have
\begin{equation}
  YZ^{3}-2ZYZ^{2}+Z^{2}YZ=(YZ^{2}-2ZYZ+Z^{2}Y)Z=0,
\end{equation}
for example. This demonstrates again that our quantization procedure results
in a unique NC hypersurface which we may then choose to represent in its
symmetrized form (corresponding to $a=1/3$ for $m=2$, and $b=c=1/4$ for $m=3$):
\begin{eqnarray}
 A_{n-1}:&&\ \ V_{A_{n-1}}(X,Y,Z):=X^2+Y^2 +Z^{n},  \notag \\
 D_{n}:&&\ \ V_{D_n}(X,Y,Z):=
  X^{2}+\frac{1}{3}\left( ZY^{2}+YZY+Y^{2}Z\right)+Z^{n-1},  \notag \\
 E_{6}:&&\ \ V_{E_6}(X,Y,Z):=X^{2}+Y^{3}+Z^{4},  \nonumber \\
 E_{7}:&&\ \ V_{E_7}(X,Y,Z):=X^{2}+Y^{3}+\frac{1}{4}\left(
  YZ^{3}+ZYZ^{2}+Z^{2}YZ+Z^{3}Y\right),  \notag \\
 E_{8}:&&\ \ V_{E_8}(X,Y,Z):=X^{2}+Y^{3}+Z^{5}. 
\label{nc}
\end{eqnarray}

Upon replacing the operators $X,Y$ and $Z$ by their differential-operator
representations given in (\ref{XYZ}), it is straightforward
to write down the corresponding holomorphic wave equations (\ref{VPsi}). 
This is discussed below. It is also stressed that, by construction,
the NC nature of the quantized $ADE$ spaces is inherited from the
ambient space $\mathbb{C}_\Theta^3$. An immediate way of seeing
this is that the non-commutativity in either case is governed by the 
same set of structure constants $\alpha,\beta,\gamma$.

\section{Wave-equation representation}

Here we list the differential-operator representations of the NC $ADE$ 
potentials outlined above:
\begin{eqnarray}
 V_{A_{n-1}}(X,Y,Z)&=&\sum_{j=0}^2\sum_{k=0}^j
  \left(\begin{array}{cc}2\\ j\end{array}\right)   
  \left(\begin{array}{cc}j\\ k\end{array}\right)  (-1)^k\left(\alpha^{j-k}\gamma^k
   x^{2-j}\partial_y^{j-k}\partial_z^k +\beta^{j-k}\alpha^k
   y^{2-j}\partial_z^{j-k}\partial_x^k\right) \nonumber\\
  &+&\sum_{j=0}^n\sum_{k=0}^j
  \left(\begin{array}{cc}n\\ j\end{array}\right)   
  \left(\begin{array}{cc}j\\ k\end{array}\right)  (-1)^k\gamma^{j-k}\beta^k
   z^{n-j}\partial_x^{j-k}\partial_y^k,  \nonumber \\
 V_{D_n}(X,Y,Z)&=&\sum_{j=0}^2\sum_{k=0}^j
  \left(\begin{array}{cc}2\\ j\end{array}\right)   
  \left(\begin{array}{cc}j\\ k\end{array}\right)  (-1)^k\left(
  \alpha^{j-k}\gamma^kx^{2-j}\partial_y^{j-k}\partial_z^k
   \right.\nonumber\\
  &&+ \left. \beta^{j-k}\alpha^ky^{2-j}\left(z+\gamma\partial_x-\beta\partial_y\right)
    \partial_z^{j-k}\partial_x^k  \right)\nonumber\\
  &+&\sum_{j=0}^{n-1}\sum_{k=0}^j
   \left(\begin{array}{cc}n-1\\ j\end{array}\right)   
   \left(\begin{array}{cc}j\\ k\end{array}\right)  (-1)^k\gamma^{j-k}\beta^k
   z^{n-1-j}\partial_x^{j-k}\partial_y^k,  \nonumber \\ 
 V_{E_6}(X,Y,Z)&=&\sum_{j=0}^2\sum_{k=0}^j
  \left(\begin{array}{cc}2\\ j\end{array}\right)   
  \left(\begin{array}{cc}j\\ k\end{array}\right)  (-1)^k\alpha^{j-k}\gamma^k
   x^{2-j}\partial_y^{j-k}\partial_z^k \nonumber\\
  &+&\sum_{j=0}^3\sum_{k=0}^j
  \left(\begin{array}{cc}3\\ j\end{array}\right)   
  \left(\begin{array}{cc}j\\ k\end{array}\right)  (-1)^k\beta^{j-k}\alpha^k
   y^{3-j}\partial_z^{j-k}\partial_x^k \nonumber\\
  &+&\sum_{j=0}^4\sum_{k=0}^j
  \left(\begin{array}{cc}4\\ j\end{array}\right)   
  \left(\begin{array}{cc}j\\ k\end{array}\right)  (-1)^k\gamma^{j-k}\beta^k
   z^{4-j}\partial_x^{j-k}\partial_y^k,  \nonumber \\
 V_{E_7}(X,Y,Z)&=&\sum_{j=0}^2\sum_{k=0}^j
  \left(\begin{array}{cc}2\\ j\end{array}\right)   
  \left(\begin{array}{cc}j\\ k\end{array}\right)  (-1)^k\alpha^{j-k}\gamma^k
   x^{2-j}\partial_y^{j-k}\partial_z^k \nonumber\\
  &+&\sum_{j=0}^3\sum_{k=0}^j
  \left(\begin{array}{cc}3\\ j\end{array}\right)   
  \left(\begin{array}{cc}j\\ k\end{array}\right)  (-1)^k\left(
   \beta^{j-k}\alpha^ky^{3-j}\partial_z^{j-k}\partial_x^k\right. \nonumber\\
  &&+\left. \gamma^{j-k}\beta^kz^{3-j}\left(y+\beta\partial_z-\alpha\partial_x\right)
   \partial_x^{j-k}\partial_y^k\right),  \nonumber \\
 V_{E_8}(X,Y,Z)&=&\sum_{j=0}^2\sum_{k=0}^j
  \left(\begin{array}{cc}2\\ j\end{array}\right)   
  \left(\begin{array}{cc}j\\ k\end{array}\right)  (-1)^k\alpha^{j-k}\gamma^k
   x^{2-j}\partial_y^{j-k}\partial_z^k \nonumber\\
  &+&\sum_{j=0}^3\sum_{k=0}^j
  \left(\begin{array}{cc}3\\ j\end{array}\right)   
  \left(\begin{array}{cc}j\\ k\end{array}\right)  (-1)^k\beta^{j-k}\alpha^k
   y^{3-j}\partial_z^{j-k}\partial_x^k \nonumber\\
  &+&\sum_{j=0}^5\sum_{k=0}^j
  \left(\begin{array}{cc}5\\ j\end{array}\right)   
  \left(\begin{array}{cc}j\\ k\end{array}\right)  (-1)^k\gamma^{j-k}\beta^k
   z^{5-j}\partial_x^{j-k}\partial_y^k .
 \label{ncdiff}
 \end{eqnarray}
The associated wave equations are defined by (\ref{VPsi})
for $\epsilon=0$. Below follows an analysis of the case $A_1$.

\subsection{The case $A_1$}

We consider
\begin{equation}
 V_{A_1}(X,Y,Z)\Psi=\left(X^2+Y^2+Z^2\right)\Psi=\epsilon\Psi
\label{A12}
\end{equation}
where $\epsilon=0$ corresponds to the NC $A_1$ geometry.
In terms of differential operators we have
\begin{eqnarray}
 X^2+Y^2+Z^2&=&x^2+y^2+z^2+2(\gamma z-\alpha y)\partial_x
  +2(\alpha x-\beta z)\partial_y+2(\beta y-\gamma x)\partial_z
  \nonumber\\
 &+&(\alpha^2+\gamma^2)\partial_x^2+(\alpha^2+\beta^2)\partial_y^2
  +(\beta^2+\gamma^2)\partial_z^2\nonumber\\
 &-&2\beta\gamma\partial_x\partial_y-2\alpha\gamma\partial_y\partial_z
 -2\alpha\beta\partial_x\partial_z.
\label{A12diff}
\end{eqnarray}
We also introduce the geometric angular momentum
\begin{equation}
 L=(L_x,L_y,L_z)=r\times\nabla,\ \ \ \ \ 
       r=(r_x,r_y,r_z)=(x,y,z),\ \ \ \ \ 
  \nabla=(\partial_x,\partial_y,\partial_z)
\label{Lrn}
\end{equation}
in terms of which the differential-operator representation may be written
\begin{eqnarray}
 V_{A_1}(X,Y,Z)&=&V_{A_1}(x,y,z)
  +2\left(\alpha L_z+\beta L_x+\gamma L_y\right)\nonumber\\
  &+&(\alpha^2+\beta^2+\gamma^2)\nabla^2
 -(\alpha\partial_z+\beta\partial_x+\gamma\partial_y) .
\label{Ldif}
\end{eqnarray}
The differential operators involved here are seen to satisfy 
the following commutator
\begin{equation}
 \left[\alpha L_z+\beta L_x+\gamma L_y,(\alpha^2+\beta^2+\gamma^2)\nabla^2
 -(\alpha\partial_z+\beta\partial_x+\gamma\partial_y)\right]=0.
\label{comm}
\end{equation}
It is therefore natural to look for an orthonormal coordinate system $(u,v,w)$
in terms of which $V_{A_1}(X,Y,Z)$ is independent of $L_w$. The two remaining
coordinates are chosen from a `symmetrical' point of view, as follows
\begin{eqnarray}
 x&=& \frac{\gamma-\alpha}{N_2}u+
  \frac{\gamma(\gamma-\beta)+\alpha(\alpha-\beta)}{N_4}v+
  \frac{\beta}{N}w,\nonumber\\
 y&=&\frac{\alpha-\beta}{N_2}u+
  \frac{\alpha(\alpha-\gamma)+\beta(\beta-\gamma)}{N_4}v+
  \frac{\gamma}{N}w,\nonumber\\
 z&=&
  \frac{\beta-\gamma}{N_2}u+
  \frac{\beta(\beta-\alpha)+\gamma(\gamma-\alpha)}{N_4}v+
  \frac{\alpha}{N}w,
\label{xyzuvw}
\end{eqnarray}
where
\begin{eqnarray}
 N_2^2&=&(\gamma-\alpha)^2+(\alpha-\beta)^2+(\beta-\gamma)^2,\nonumber\\
 N_4^2&=&(\gamma(\gamma-\beta)+\alpha(\alpha-\beta))^2
  +(\alpha(\alpha-\gamma)+\beta(\beta-\gamma))^2
  +(\beta(\beta-\alpha)+\gamma(\gamma-\alpha))^2,\nonumber\\
 N^2&=&\alpha^2+\beta^2+\gamma^2.
\label{NNN}
\end{eqnarray}
These normalization constants are seen to be related according to
\begin{equation}
 N_4^2=N_2^2N^2
\label{N2}
\end{equation}
After some somewhat tedious computations one finds the following
remarkable simplification
\begin{equation}
 V_{A_1}(X,Y,Z)=u^2+v^2+w^2+N^2(\partial_u^2+\partial_v^2).
\label{Vuvw}
\end{equation}
The simplicity of this expression is due to the change of coordinates
(\ref{xyzuvw}), exploiting the symmetries of the original differential operator
(\ref{A12diff}). In either form, the differential operator represents the NC elevation
of the polynomial $V_{A_1}(x,y,z)$. It thus appears in the 
reduction
of $\mathbb{C}_\Theta^3$ to the NC hypersurface defined by 
$V_{A_1}(X,Y,Z)=0$. Here we are interested in
solving the differential equation (\ref{A12}) using (\ref{Vuvw}).

The NC system may now be studied by representing the orthonormal
coordinates $(u,v,w)$ in cylindrical coordinates 
\begin{equation}
 u=\rho\cos(\theta),\ \ \ \ \ \ \ v=\rho\sin(\theta),\ \ \ \ \ \ \ 
  w=w,\ \ \ \ \ \ \ \partial_u^2+\partial_v^2
  =\partial_\rho^2+\frac{1}{\rho}\partial_\rho
  +\frac{1}{\rho^2}\partial_\theta^2,
\label{uvrhotheta}
\end{equation}
in which case the differential equation (\ref{A12}) reads
\begin{equation}
 \left(\rho^2+w^2+N^2
  \left(\partial_\rho^2+\frac{1}{\rho}\partial_\rho
  +\frac{1}{\rho^2}\partial_\theta^2\right)\right)\Psi(\rho,\theta;w)
   =\epsilon\Psi(\rho,\theta;w).
\label{diffN}
\end{equation}
Since this equation does not involve derivatives with respect to $w$,
we consider the following simple separation of variables
\begin{equation}
 \Psi(\rho,\theta;w)=R_w(\rho)\Upsilon(\theta)
\label{sep}
\end{equation} 
with corresponding differential equations
\begin{eqnarray}
 \left(N^2\rho^2\frac{d^2}{d\rho^2}+
  N^2\rho\frac{d}{d\rho}+\rho^4+(w^2-\epsilon)\rho^2
  -\mu^2\right)R_{w,\mu}(\rho)&=&0,\nonumber\\
 \left(\frac{d^2}{d\theta^2}+\frac{\mu^2}{N^2}
  \right)\Upsilon_\mu(\theta)&=&0.
\label{sepdiff}
\end{eqnarray}

The angular equation is the well-known differential equation
for the harmonic oscillator. It has two linearly independent solutions:
\begin{equation}
 \Upsilon_\mu^+(\theta)=e^{i\mu\theta/N},\ \ \ \ \ \ \ \ \ 
 \Upsilon_\mu^-(\theta)=e^{-i\mu\theta/N}.
\label{har}
\end{equation}

After making the substitution
\begin{equation}
 R(\rho)=\frac{1}{\rho}Q(i\rho^2/N)
\label{RW}
\end{equation}
for the radial function, we find the differential equation
\begin{equation}
 \left(\frac{d^2}{d(\frac{i\rho^2}{N})^2}
 +\left(-\frac{1}{4}+\frac{i(\epsilon-w^2)/(4N)}{i\rho^2/N}+\frac{\frac{1}{4}
 -\left(\frac{\mu}{2N}\right)^2}{\left(i\rho^2/N\right)^2}\right)\right)Q(i\rho^2/N)=0.
\label{Wdiff}
\end{equation}
This is recognized as the Whittaker differential equation whose
two linearly independent solutions may be represented by
\begin{equation}
 M_{\frac{i(\epsilon-w^2)}{4N},\frac{\mu}{2N}}(i\rho^2/N)\ \ \ {\rm and}\ \ \ 
 M_{\frac{i(\epsilon-w^2)}{4N},-\frac{\mu}{2N}}(i\rho^2/N)
\end{equation}
or in terms of Whittaker's function by
\begin{equation}
 W_{\frac{i(\epsilon-w^2)}{4N},\frac{\mu}{2N}}(i\rho^2/N)\ \ \ {\rm and}\ \ \ 
 W_{-\frac{i(\epsilon-w^2)}{4N},\frac{\mu}{2N}}(-i\rho^2/N).
\label{solw}
\end{equation}
To see this, it is recalled \cite{GR}
that the Whittaker differential equation is given by
\begin{equation}
 \frac{d^2W(z)}{dz^2}+\left(-\frac{1}{4}+\frac{\lambda}{z}+
  \frac{\frac{1}{4}-\kappa^2}{z^2}\right)W(z)=0,
\label{W}
\end{equation}
and that it has the two linearly independent solutions
\begin{eqnarray}
 M_{\lambda,\kappa}(z)&=&z^{\kappa+\frac{1}{2}}e^{-z/2}
  \Phi(\kappa-\lambda+\frac{1}{2},2\kappa+1;z),\nonumber\\
 M_{\lambda,-\kappa}(z)&=&z^{-\kappa+\frac{1}{2}}e^{-z/2}
  \Phi(-\kappa-\lambda+\frac{1}{2},-2\kappa+1;z).
\label{WM}
\end{eqnarray}
Here $\Phi(\nu,\tau;z)$ denotes the confluent hypergeometric 
function sometimes written ${}_1F_1(\nu;\tau;z)$.
Whittaker's function provides solutions suitable for $2\kappa$
integer, and are defined by
\begin{equation}
 W_{\lambda,\kappa}(z)=\frac{\Gamma(-2\kappa)}{\Gamma(\frac{1}{2}
  -\kappa-\lambda)}M_{\lambda,\kappa}(z)
  +\frac{\Gamma(2\kappa)}{\Gamma(\frac{1}{2}
  +\kappa-\lambda)}M_{\lambda,-\kappa}(z).
\label{WW}
\end{equation}
Two linearly independent solutions to (\ref{W}) of this kind are given by
$W_{\lambda,\kappa}(z)$ and $W_{-\lambda,\kappa}(-z)$.

Since the spectrum of solutions to the differential equation
(\ref{A12}) is given in terms of the harmonic oscillator and
solutions to the Whittaker differential equation, the involved parameters
are not constrained by quantization conditions in the usual sense.
Rather, the quantization conditions manifest themselves 
in the {\em form} of the spectrum which in this case is comprised
of a combination of (\ref{har}) and (\ref{solw}).

Now that we have the solution to (\ref{A12}) for all $\epsilon$, it is natural
to study the limit $\epsilon\rightarrow0$.
In the notation of (\ref{W}), the only dependence on $\epsilon$ is through
$\lambda$. The Whittaker differential equation and its solution are
well defined for all $\lambda$, so we may conclude that the differential
equation (\ref{Wdiff}) is well defined for all $\epsilon$.
The original, classical $A_1$ geometry, on the other hand,
is singular and corresponds to the aforementioned limit:
\begin{equation}
 V_{A_1}(x,y,z)=\epsilon\rightarrow0.
\label{Ve0}
\end{equation}
A merit of our quantization procedure is thus that the singularity of
the classical geometry has been resolved. This NC elevation therefore
offers an alternative to the more conventional resolutions.

Due to the interpretation that the differential operator (\ref{A12diff})
represents the NC elevation of the singular K3 surface
$x^2+y^2+z^2=0$,
we find it natural to attribute all the solutions to (\ref{A12}) 
to the `moduli space' of the associated NC geometry.
That is, the spectrum of wave functions solving (\ref{A12}),
or more generally (\ref{VPsi}), for $\epsilon=0$ is interpreted as
the moduli space of the NC geometry.
Since the latter is represented by a partial differential equation,
we see that its possible boundary conditions correspond 
to constraints imposed on the moduli space.
A detailed analysis of this link between boundary conditions
and constraint equations is beyond the scope of
the present work.

In the limit of zero non commutativity, i.e., $\alpha=\beta=\gamma=0$,
the change of coordinates (\ref{xyzuvw}) is singular. This is in accordance
with the fact that the differential equation (\ref{A12}) merely reduces
to its classical counterpart 
\begin{equation}
 (x^2+y^2+z^2)\Psi=\epsilon\Psi .
\label{classA1}
\end{equation}
On the hypersurface (\ref{Veps}) for $V=V_{A_1}$, every 
complex function solves (\ref{classA1}).
This means that the `classical moduli space' may be identified
with the set of holomorphic functions.

\section{Discussion}

We have developed a new and essentially non-geometric approach to
NC Calabi-Yau manifolds, based on ideas from quantum mechanics. 
Our focus has been on the singular $ADE$ geometries. The 
polynomial equations defining these classical 
$ADE$ geometries are replaced by differential equations in which
the original singularities are (presumably) absent. 
The moduli space associated to 
such an NC geometry is then interpreted as the spectrum of solutions to the
corresponding wave equation. We have analyzed in detail the NC elevation
of the $A_1$ geometry and found that it is described in part by
the Whittaker differential equation.
We intend to discuss elsewhere the extension of this explicit
study to the other $ADE$ geometries \cite{work}.

Our approach is adaptable to a broad variety of geometries
whose NC elevations may then be represented by differential
wave equations. The extension from complex to real variables
is straightforward, as is the extension to other dimensions
than two complex ones. 
We anticipate that the NC elevations of singular geometries
in general will be non singular as in the case of $A_1$ discussed above.
This will be addressed elsewhere \cite{work} where we also intend
to discuss the implementation of boundary conditions alluded to above. 

In order to put the analogy between our construction and quantum mechanics
to a `physical test', one could examine the `dual' descriptions of the NC elevations.
That is, on one hand
we have introduced the NC elevations as `hypersurfaces'
in an NC ambient space, while on the other hand we are
representing them as differential operators resulting in
some wave equations. In quantum mechanics, this corresponds
to an operator description versus a description in terms of wave functions.
It would therefore be of interest to try to extract 
information on an NC elevation based on both its dual descriptions. 
We believe that these complimentary approaches deserve 
to be studied further.

\begin{acknowledgement}
Saidi would like to thank Protars III program, D12/25/CNRST, for support.
\end{acknowledgement}

\end{document}